\documentclass[reqno,10pt]{amsart}
\usepackage{hyperref}
\usepackage{graphicx}
\usepackage{slashed}
\usepackage{amscd}
\usepackage{tensor}
\usepackage{amssymb}
\usepackage{esint}
\usepackage{enumitem}
\usepackage{accents}
\usepackage[dvipsnames]{xcolor}
\usepackage[utf8]{inputenc}
\usepackage[off]{auto-pst-pdf}
\usepackage{pst-grad}
\usepackage{pst-plot}
\usepackage[mathscr]{eucal}
\usepackage{tagging}
\droptag{NotReady}

\numberwithin{equation}{section}
\allowdisplaybreaks[1]

\newtheorem{Def}{Definition}

\newtheorem{Lemma}{Lemma}
\newtheorem{Impl}{Implication}

\newtheorem{Example}[Def]{Example}

\newtheorem{Assumption}{Assumption}

\newcommand{\beq}{\begin{equation}}
\newcommand{\eeq}{\end{equation}}
\newcommand{\Proof}{\begin{proof}}
\newcommand{\QED}{\end{proof} \noindent}
\newcommand{\QEDrem}{\ \hfill $\Diamond$}

\newcommand{\C}{\mathcal{C}}
\newcommand{\CTM}{\C_{\textrm{TM}}}
\newcommand{\CIM}{\C_{\textrm{Imm}}}
\newcommand{\Prog}{\mathcal P\!\,r\!\,o\!\,g}
\newcommand{\cref}{c_{\textrm{ref}}}

\newcommand{\Sys}{\ensuremath \textit{Sys}}

\DeclareFontFamily{OT1}{rsfso}{}
\DeclareFontShape{OT1}{rsfso}{m}{n}{ <-7> rsfso5 <7-10> rsfso7 <10-> rsfso10}{}
\DeclareMathAlphabet{\mycal}{OT1}{rsfso}{m}{n}

\newcommand{\iD}{\raisebox{0.125em}{\tiny $\blacktriangleright$}}

\newcommand{\bei}{\begin{itemize}[label=$\circ$,itemsep=.5em,leftmargin=*]}
\newcommand{\beii}{\begin{itemize}[label=$\rightarrow$,itemsep=.5em,topsep=.5em,leftmargin=*]}
\newcommand{\eni}{\end{itemize}}
\newcommand{\enii}{\end{itemize}}

\definecolor{darkblue}{RGB}{0,91,163}

\usepackage{phaistos}

\usepackage{mdframed} 

\newcommand{\ql}[1]{[\rotatebox{90}{\scalebox{0.25}{later}}]}

\newcommand{\cbb}[1]{   
\begin{itemize}[label=\iD,topsep=.5em,itemsep=.5em,leftmargin=*]
\item \textbf{#1}
    \begin{itemize}[label=-,itemsep=.2em,topsep=.2em,leftmargin=*]
}
\newcommand{\cbe}{      
    \end{itemize}
\end{itemize}
}

\usepackage{multicol}
\usepackage{endnotes}
\usepackage{apacite}

\usepackage{mdframed}
\definecolor{lightblue}{RGB}{225,244,249}			
\definecolor{lightgreen}{RGB}{218,247,232}          
\mdtheorem[style=Box,backgroundcolor=Bsp]{example}{Example}[Def]
\theoremstyle{plain}
\newmdenv[linewidth=0,needspace=130pt,skipabove=5pt,skipbelow=5pt,innertopmargin=10pt,
splittopskip=30pt,splitbottomskip=18pt,innerbottommargin=10pt,frametitleaboveskip=0pt,backgroundcolor=lightblue]{quick-example}
\newmdenv[linewidth=0,skipabove=5pt,skipbelow=5pt,innertopmargin=10pt,
splittopskip=30pt,splitbottomskip=18pt,innerbottommargin=10pt,frametitleaboveskip=0pt,backgroundcolor=lightblue]{defi}
\newmdenv[linewidth=0,skipabove=5pt,skipbelow=5pt,innertopmargin=10pt,
splittopskip=30pt,splitbottomskip=18pt,innerbottommargin=10pt,frametitleaboveskip=0pt,backgroundcolor=lightblue]{terminology}
\newmdenv[linewidth=0,skipabove=5pt,skipbelow=5pt,innertopmargin=10pt,
splittopskip=30pt,splitbottomskip=18pt,innerbottommargin=10pt,frametitleaboveskip=0pt,backgroundcolor=lightgreen]{reality}

\usepackage{cases}

\makeatletter
\def\section{\@startsection{section}{1}%
  \z@{.7\linespacing\@plus\linespacing}{.5\linespacing}%
  {\Large\scshape\centering}}
\makeatother

\usepackage{soul}       

\begin{document}


\begin{center}
\Huge Consciousness qua Mortal Computation
\end{center}

\vspace{1cm}

\centerline{\Large Johannes Kleiner$^{1,2,3,4}$}
\vspace*{.3cm}
\centerline{$^1$Munich Center for Mathematical Philosophy, Ludwig-Maximilians-Universität München}
\centerline{$^2$Graduate School of Systemic Neurosciences, Ludwig-Maximilians-Universität München}
\centerline{$^3$Institute for Psychology, University of Bamberg}
\centerline{$^4$Association for Mathematical Consciousness Science}

\vspace{.7cm}

\begin{quote}
\textsc{Abstract.}
Computational functionalism posits that consciousness is a computation. Here we show, perhaps surprisingly, that it cannot be a Turing computation. Rather, computational functionalism implies that consciousness is a novel type of computation that has recently been proposed by Geoffrey Hinton, called mortal computation.
\end{quote}

\vspace{.7cm}

\begin{multicols}{2}


\section{Introduction}

A fundamental tenet of general purpose digital computing is that software is separated from hardware, so that the same program or algorithm can be run on any suitable system. 
This tenet is about to be broken. Contemporary developments in Artificial Intelligence~(AI) and AI chip production have led to the identification of a novel concept of general purpose computing, called \emph{mortal computation}~\cite{hinton2022forward}. This concept draws a line between the type of computations that contemporary processing units do, and the type of computations that brains and other biological organisms carry out.

Computational functionalism, first defined by~\citeA{putnam1967psychological}, posits, in a nutshell, that consciousness is a computation.
This view has gained popularity again in light of the staggering achievements in AI development in recent years. AI models are computations,
so if computational functionalism is true, AI models can---and, depending on the nature of the computation that consciousness is, will---become
conscious~\cite{butlin2023consciousness}.

Here we show that computational functionalism is not indifferent with respect to the type of computation that consciousness is. 
We show that if there is any organism that is capable of conscious experiences, but which cannot be programmed---for example, non-human animals; cf. Assumption~\ref{asmpt:tildeS}---, then computational functionalism implies that consciousness is a mortal computation. To establish this result, we make use of a differential definition of mortal computation, as well as general facts about the relation
between programs, Turing computation and immortal computation.

Our result challenges the usual understanding of computational functionalism, which is centered around Turing-like models of computation. If our result holds true,
consciousness cannot, according to computational functionalism, be a Turing computation or programmed. Yet, contemporary AI systems and programs are Turing computations. Therefore,
this result speaks against the possibility of AI consciousness (though it does not aim to settle the issue due to questions of realization, cf. Section~\ref{sec:conclusion}). 

The underlying perspective of this paper is that the discovery of mortal computation by~\citeA{hinton2022forward} may well be a first step towards understanding of a whole new paradigm of computation, potentially as consequential as the Turing-Church-G\"odel-Herbrand paradigm of computation of the past nine decades~\cite{goedel1934undecidable,church1936unsolvable,turing1936computable}.%
    \endnote{The question of whether a computation is a Turing \emph{computation} is different from questions regarding Turing~\emph{computability}.
    The former concerns the nature of computations. For example, the question of whether neural computations are Turing computations~\cite{piccinini2020neurocognitive}.
    The latter concerns functions, in the mathematical sense of the term, that map natural numbers to natural numbers, and asks whether their value can be computed by a Turing machine. A function is Turing-computable iff there is a Turing computation (meaning: an abstract mathematical model of a Turing machine) that halts on all numbers for which the function is defined, and does not halt when provided with numbers for which the function is not defined. This is the case iff the function is $\lambda$-computable \cite{church1936unsolvable,turing1937computability} or general recursive \cite{goedel1934undecidable,kleene1936lambda}. The definition of Turing-computability of functions leaves open what the computation is that implements the function, which is what this paper is concerned with.
    }

\section{Mortal Computation}

The notion of mortal computation was identified and coined by~\citeA[Sec.\,9]{hinton2022forward},
who describes a learning task that makes use of unknown properties of hardware that vary across systems, such as variations in the connectivity of a system, or variations in non-linear processes in a system.
As a result, the parameter values that define the learned computation ``are only useful for that specific hardware instance, so the computation they perform is mortal: it dies with the hardware''~\cite[p.\,13]{hinton2022forward}.
The general computing paradigm of the past nine decades,
in contrast, implies
that a computation is largely independent of the hardware on which it is run:
``[T]he same program or the same set of weights can be
run on a different physical copy of the hardware. This makes the knowledge contained in the program or the weights immortal: The knowledge does not die when the hardware dies''~\cite[p.\,13]{hinton2022forward}. 

There is, at this early stage, no constructive definition of mortal computation,%
    \endnote{
    There are two ways of reading~\citeA[Sec.\,9]{hinton2022forward}. On a deflationary reading,
    a mortal computation is simply a Turing computation that is not known in its entirety to an outside programmer. Call this
    the epistemic reading of mortal computation. It is suggested by Hinton's emphasis of ``large and unknown variations
    in the connectivity'' (ibid.). On a different reading, a mortal computation is a computation that fundamentally
    transcends some of the constraints of Turing computation, for example the existence of an immutable tape for purposes other than read and write actions, or
    the existence of a transition function that is Markov, as suggested by Hinton's emphasis on ``non-linearities of different instances of hardware'' (ibid.). Call this the ontic reading of mortal computation.
    On the ontic reading, the state of affairs of the hardware is partially unknown to the computation itself. The computation may have to deal with,
    and make use of, non-Turing properties of the hardware. 
    Both interpretations are compatible with~\eqref{def:ImmSubsetTM}.
    }
but we may consider a differential definition, that helps us distinguish mortal computations in virtue of what they are not. To provide such definition, denote by $\C$ the class of all computations. $\C$~comprises all Turing computations, which we will denote by $\CTM$ in what follows,
as well as other notions of computation, for example, non-deterministic Turing computations, neural computations, analogue computations and the yet-to-be-understood mortal computations.

The core intuition behind immortal computation is
``that the software should be separable from the hardware''~\cite[p.\,13]{hinton2022forward}.
In practice---in central processing units (CPUs), graphics processing units (GPUs), tensor processing units (TPUs), or data processing units (DPUs)---this separation is enabled by a processing unit's \emph{Instruction Set Architecture} (ISA). An ISA contains specifications of all computations that the processing unit can carry out, and it is with respect to these specifications that programs, operating systems and compilers are defined. To run a program is to run machine code that specifies which of the ISA's computations are to be carried out in which order (call this concatenation) and how the results of computations are to be used by other computations (call this combination).
Differences among processing units' performance, design, size, etc., are differences in an ISA's \emph{implementation}. The ISA exists to ensure binary-code compatibility of software despite such differences; it is the boundary between software and hardware.

The computations defined by an ISA constitute a reference relative to which software is defined, and which a class of hardware implements. It ensures that a program can run on different physical copies of the same type of hardware. Computation is immortal precisely because it is defined with respect to such reference. We can formalize this requirement as follows.

\begin{Def}\em\label{def:immortal}
A computation $c \in \C$ is \emph{immortal} iff there is a class of reference computations $\cref \subset \C$
such that $c$ is a concatenation and combination of these reference computations. A computation $c$ is \emph{mortal} iff it is not immortal.
\end{Def}

We will denote the class of immortal computations by $\CIM$.
Immortal computations are meant to be a subclass of Turing computations, so that we have 
\begin{align}\label{def:ImmSubsetTM}
    \CIM \subset \CTM \:.
\end{align}

Because an immortal computation $c$ is a concatenation and combination of reference computations, every system that can realize an immortal computation $c$ must be able to realize its reference computations $\cref$. This is the only implication of Definition~\ref{def:immortal} we will make use of in what follows. To explicate this implication formally, we denote by $\Sys$ the class of all systems. This class includes, for example, all CPUs, GPUs, TPUs, and DPUs in use today, as well as all biological organisms. Furthermore, we denote by $\C(S)$ all computations that a system $S \in \Sys$ can realize or implement. Using such formalism is of advantage because it can be applied to any account of implementation of a computation~\cite{piccinini2015physical}. The essential implication of the previous definition then reads as follows.

\begin{Impl}\em
If $c \in \C$ is immortal, then there is a class $\cref \subset \C$ such that for all $S \in \Sys$
\begin{align}
c \in \C(S) &\Rightarrow \cref \subset \C(S) \label{impl:immortal1} \:. 
\end{align}
\end{Impl}

An important class of immortal computations are computations specified by writing a program in some programming language; computations that are coded, that is, and compiled to run on CPUs, GPUs, TPUs or DPUs. We will simply refer to these computations as `programs'. Programs are immortal because they are defined with respect to some programming language that in turn is defined, via its compiler, with respect to one or more ISAs.

We denote the class of computations that can be coded with any of the existing programming languages by $\Prog$. Because programs are immortal, we have
\begin{align}\label{eq:progImm}
\Prog \subset \CIM \:.   
\end{align}

\section{Computational Functionalism}\label{sec:CF}

Computational functionalism was introduced by~\citeA{putnam1967psychological} as the following set of assumptions.
    \begin{enumerate}[label=\arabic*.,leftmargin=2em,rightmargin=1em,itemsep=.2em,topsep=.2em]
        \item ``All organisms capable of feeling pain are Probabilistic Automata.
        \item Every organism capable of feeling pain possesses at least one Description of a certain kind (i.e., being capable of feeling pain \emph{is} possessing an appropriate kind of Functional Organization).
        \item No organism capable of feeling pain possesses a decomposition into parts which separately possess Descriptions of the kind referred to in 2.
        \item For every Description of the kind referred to in 2, there exists a subset of the sensory inputs such that an organism with that Description is in pain when and only when some of its sensory inputs are in that subset.''~\cite[p.\,434]{putnam1967psychological,putnam1975nature}
    \end{enumerate}

In giving this definition, Putnam equates Probabilistic Automata with \emph{descriptions} of a system;
``[t]he Machine Table mentioned in the Description will then be called the
Functional Organization of [a system] $S$ relative to that Description''~(ibid.).

The understanding of computation has evolved substantially since~\citeA{putnam1967psychological}, cf. e.g. \cite{piccinini2015physical}. To connect Putnam's definition to computation as presently understood, and to do justice of it being a definition of computational functionalism, we reformulate Condition~2 in abstract terms, making use of the set $\C(S)$ of computations that a system can realize, which we have introduced above.
It is clear from the context of Putnam's definition that it is to apply to all systems, not just organisms in a narrow sense. 
Denoting the experience of ``feeling pain'' by $e$, and the class of systems capable of having this experience by $\Sys_e$, we may hence read Putnam's Condition~2 as follows.

\begin{Impl}\em\label{impl:CF}
    Computational functionalism implies that
    there is at least one computation $c^\ast \in \C$  such that, for all $S \in \Sys$,
    \begin{equation*}
        S \in \Sys_e \ \Rightarrow \  c^\ast \in \C(S) \:.
    \end{equation*}   
\end{Impl}

In Putnam's terms, being capable of realizing $c^\ast$ is being capable of experiencing pain.
Modulo details of sensory input referred to in Putnam's Condition~4, we may say that experiencing pain \emph{is} realizing a computation $c^\ast$, or,
in more simple terms yet, that the experience of pain is $c^\ast$. Nothing hinges on these terminological shortcuts, though. 
Putnam's conditions must hold as well for experiences other than pain, but may be realized by different computations in each case.

\section{Programs}

We have denoted the class of computations that can be coded with any existing programming language by $\Prog$.
We now define $\Sys_0$ to denote the class of systems that can run such programs. Because programs are defined relative
to Instruction Set Architectures (ISA) of the underlying programming language, $\Sys_0$ is the class of systems that can realize ISAs of existing programming languages.
Denoting, as above, an ISA of a program $c \in \Prog$ by~$\cref$, this class is defined as 
\begin{align}\label{def:Sys0}
\begin{split}
    \Sys_0 = \{ S \in \Sys \ | \ & \cref \subset \C(S) \textrm{ for at} \\
    &\textrm{least one } c \in \Prog \} \:.
\end{split}
\end{align}

The class $\Sys_0$ comprises all desktop and laptop computers, mobile devices, workstations, servers and supercomputers. It comprises anything that can run
any Instruction Set Architecture of any existing compiler or programming language. But it does not comprise animals, or other organisms, that cannot be programmed---it does not contain organisms which are incapable of operating non-trivial logic as required by ISAs, that is. If any such animal or orgamism is conscious, the following assumption holds true.

\begin{Assumption}\em\label{asmpt:tildeS}
There is a system $S \not\in \Sys_0$ that is capable of conscious experience $e$.
\end{Assumption}

In what follows, we assume computational functionalism (viz. Implication~\ref{impl:CF})
and Assumption~\ref{asmpt:tildeS}. The following lemma shows that if this is the case, the computation $c^\ast$ is not among all programs.

\begin{Lemma}\em\label{lem:notProg}
    $c^\ast \not\in \Prog$.
\end{Lemma}

\Proof
Assume $c^\ast \in \Prog$ and let $\tilde S$ denote the system in Assumption~\ref{asmpt:tildeS}.
Because $\tilde S \not\in \Sys_0$,
it follows that $\cref^\ast \not\subset \C(\tilde S)$.
But because $\tilde S \in \Sys_e$, Implication~\ref{impl:CF}
implies that $c^\ast \in \C(\tilde S)$.
This violates \eqref{impl:immortal1}, so that
$c^\ast$ cannot be immortal. But
all programs are immortal (cf. \eqref{eq:progImm}). Hence we have arrived at a contradiction. 
It follows that $c^\ast \not\in \Prog$.
\QED

\section{Turing Computation}

Next, we consider Turing computations. A computation is a Turing computation
iff it can be realized by (the abstract mathematical model of) a Turing machine.%

Some (in fact, most) contemporary programming languages are Turing complete: they can be used to simulate universal Turing machines, meaning that they can be used to implement any Turing computation. For any Turing computation, one can write at least one program that realizes this computation, and running this program instantiates the Turing computation.
This implies that
\begin{align}\label{eq:CTMprog}
    \CTM \subset \Prog \:.
\end{align}

As a consequence, we have the following lemma, which shows that the computation $c^\ast$ put forward by Computational Functioanlism
is not a Turing computation.

\begin{Lemma}\label{lem:notCtm}
$c^\ast \not\in \CTM$ \:. 
\end{Lemma}

\Proof
Follows from Lemma~\ref{lem:notProg} and~\eqref{eq:CTMprog}.
\QED

\section{Immortal Computation}

The next lemma shows that consciousness is a mortal computation.

\begin{Lemma}
   $c^\ast \not\in \CIM$.
\end{Lemma}

\Proof
Lemma~\ref{lem:notCtm} states that
$c^\ast \not\in \CTM$.
Because of~\eqref{def:ImmSubsetTM}, we furthermore have
$\CIM \subset \CTM$. Therefore, it follows that
$c^\ast \not\in \CIM$.
\QED

\section{Conclusion}\label{sec:conclusion}

We have shown that computational functionalism implies that consciousness is a mortal computation, and that consciousness cannot be a program or Turing computation.
We hope that this result contributes to the understanding of computational functionalism and its implications, including questions of AI consciousness,
and that it highlights mortal computation as a potential concept of interest with respect to question of the mind.

Because all contemporary Artificial Intelligence (AI) is immortal computation, the results provide initial reason to believe that no current or near-future AI can be conscious. Only artificial systems that employ mortal computations can instantiate consciousness. The results presented here do not, however, prove this to be the case.
That is because it might be the case that consciousness, despite being a mortal computation, can be realized by immortal computations. Whether this is a viable option,
and what precisely it means to realize or implement a mortal computation, depends on details of the notion of mortal computation that are to be developed in future research.
Because mortal computation is not Turing computation, the possibility of such realization might bear various difficulties, in case of which strong implications for synthetic phenomenology would follow.

\subsection*{Acknowledgments}
I would like to thank Hanna Tolle, Tim Ludwig, Wanja Wiese, Justin Sampson, Zhuoqiao Yin, Jonathan Mason, and David Chalmers for valuable discussions on mortal computation, as well as the organisers and participants of the \emph{C3: Complexity, Computers, and Consciousness} workshop at the \emph{Institute of Physics} for valuable feedback on earlier ideas.
I would like to thank the NYU Center for Mind, Brain, and Consciousness for hosting me while working on this manuscript.

\theendnotes

\renewcommand\bibliographytypesize{\small}

\end{multicols}
\end{document}